\documentclass{ws-ijmpcs}

\begin{document}

\markboth{U. Barres de Almeida, P. Giommi \& C.H. Brandt}
{Brazilian Science Data Center}

%
\catchline{}{}{}{}{}
%

\title{The Brazilian Science Data Center (BSDC)}

\author{Ulisses Barres de Almeida and Benno Bodmann}

\address{Centro Brasileiro de Pesquisas Físicas (CBPF), Rua Dr. Xavier Sigaud 150\\
Rio de Janeiro, RJ 22610-010,
Brazil\\
ulisses@cbpf.br; benno.bodmann@gmx.de}

\author{Paolo Giommi}

\address{Agenzia Spaziale Italiana (ASDC), Via del Politecnico snc,
Rome, 00133, Italy\\
paolo.giommi@asdc.asi.it}

\author{Carlos H. Brandt}

\address{Universit\`{a} di Roma "La Sapienza" (ICRANet), P.le Aldo Moro 5
Rome, 00185, Italy\\
carlos.brandt@asdc.asi.it}

\maketitle


\begin{abstract}

Astrophysics and Space Science are becoming increasingly characterised by what is now known as ''big data", the bottlenecks for progress partly shifting from data acquisition to ''data mining". Truth is that the amount and rate of data accumulation in many fields already surpasses the local capabilities for its processing and exploitation, and the efficient conversion of scientific data into knowledge is everywhere a challenge. The result is that, to a large extent, isolated data archives risk being progressively likened to "data graveyards", where the information stored is not reused for scientific work. 

Responsible and efficient use of these large datasets means democratising access and extracting the most science possible from it, which in turn signifies improving data accessibility and integration. Improving data processing capabilities is another important issue specific to researchers and computer scientists of each field. The project presented here wishes to exploit the enormous potential opened up by information technology at our age to advance a model for a science data center in astronomy which aims to expand data accessibility and integration to the largest possible extent and with the greatest efficiency for scientific and educational use. Greater access to data means more people producing and benefiting from information, whereas larger integration of related data from different origins means a greater research potential and increased scientific impact.

The project of the BSDC is preoccupied, primarily, with providing tools and solutions for the Brazilian astronomical community. It nevertheless capitalizes on extensive international experience, and is developed in cooperation with the ASI Science Data Center (ASDC), from the Italian Space Agency, granting it an essential ingredient of internationalisation. The BSDC is Virtual Observatory-compliant and supportive of the "Open Universe", a global initiative in discussion under the auspices of the United Nations. 

\end{abstract}


\ccode{PACS numbers: 95.75.Pq, 95.75.Wx, 95.80+p}

\section{Motivation: a view from blazar astrophysics}	

We would like to start by providing some concrete motivation for the development of a large-scale, globally integrated astronomical database such as proposed here. It is widely recognised that Astrophysics and Space Science are prime examples of international collaborative research. Today's observation of the cosmos involve expensive satellite and often large ground-based facilities, and special geographic conditions are required for the installation of modern telescopes. Individual facilites, while certainly valuable, are of limited utility when working alone, as the combination of data from a number of instruments, forming what astronomers call a multi-frequency, or multi-messenger\footnote{That is, combining, other than usual electromagnetic carriers, information derived from neutrino, cosmic-rays and even gravitational-wave observations.} view of the celestial sources, is nowadays mandatory for progress in research. In many areas, the unfolding of the cosmic drama happens at timescales that range from the extremely short (even for human standards), requiring fast integration and communication between facilities around the globe, to the extremely long, demanding the build up of extensive catalogues of source populations which in turn requires sustained observational efforts over decades. The study of blazars, for which we are developing specific analysis and data handling software tools, are perhaps the simplest, and at the same time one of the richest examples to illustrate the concept and potential of a data center like the BSDC, and will be used as a example in the following. The concept of the BSDC, of a web-based, "science-ready" data center, is readily extensible to other fields, not only of astrophysics, but of science in general, and could be of broader interest and usefulness to other communities.\\

Almost fifty years ago it was suggested, and now it is commonly accepted, that supermassive black holes (SMBHs), with masses between 10$^6$-10$^{10}$ M$_\odot$, are present at the nucleus of every Galaxy with stellar bulges.\cite{Cel99} A small fraction of these SMBHs are being fed with sufficient quantities of gas so that an accretion disk is formed around them. Radiative emission from these systems is comparable to the gross starlight output of the entire host galaxy, thanks to the large efficiency of the accretion process in converting matter to radiation, this being the fundamental mechanism by which the so-called active galactic nuclei, or AGN, shine.\cite{Ant12} For some AGN, the accretion process may be accompanied by the bulk acceleration of particles at ultra-relativistic energies into collimated jets of plasma extending for thousands of light-years away from the central engine.\cite{Fer88} AGN and their associated jets are the most luminous persistent sources of radiation in the Universe, emitting over the entire electromagnetic spectrum. Their action is fundamental within the models of evolution of galaxies and large-scale structure in the Universe.

\begin{figure}[pb]
\centerline{\includegraphics[width=13.cm]{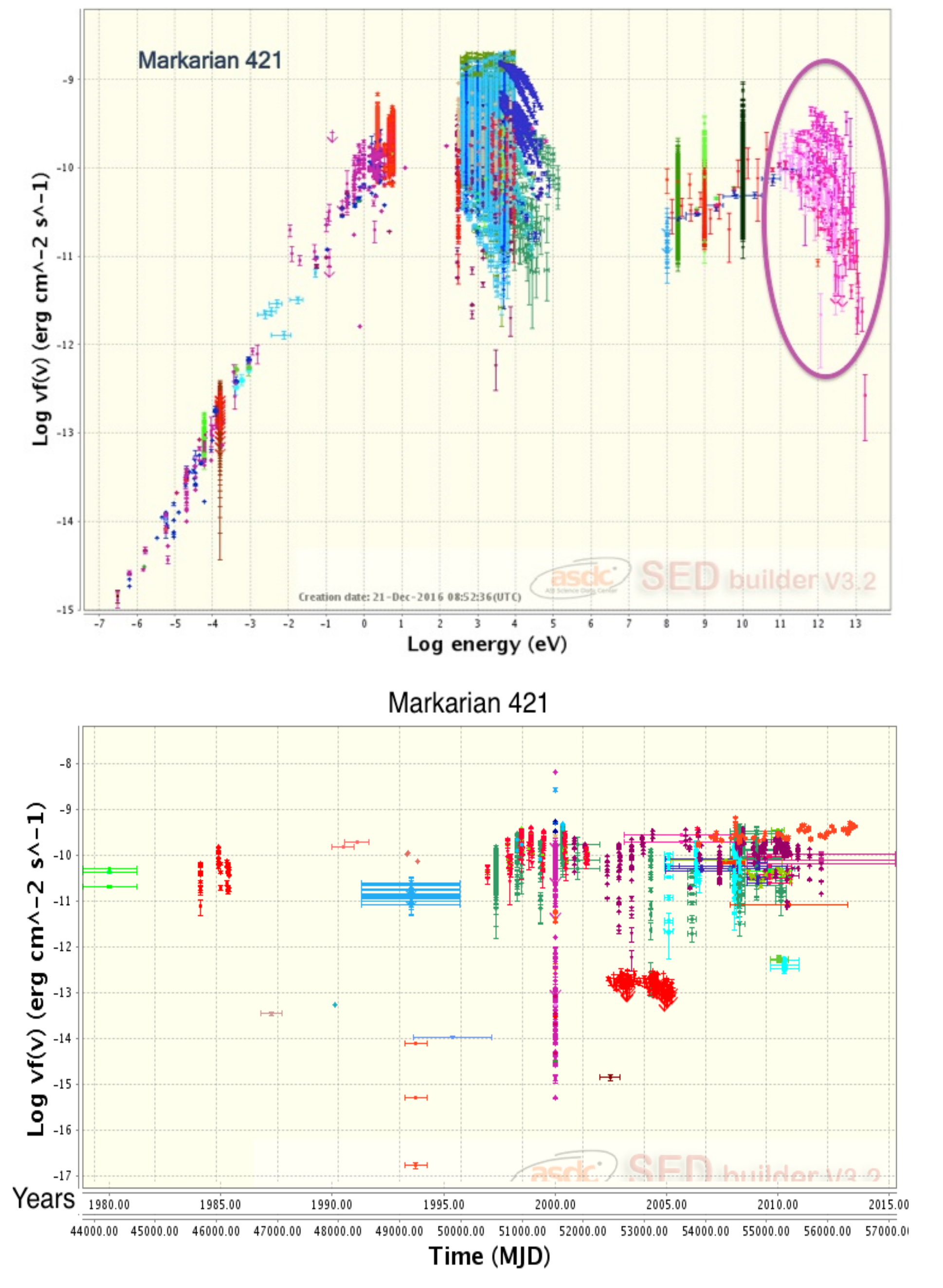}}
\vspace*{8pt}
\caption{Example of a data set. (Top panel) The spectral energy distribution (SED) for the prototypical blazar Mkn 421, plotted in the traditional energy flux representation ($\nu F_{\nu}$ vs. $\nu$), and showing the extent of the source's emission from radio to TeV gamma-rays. The marked VHE gamma-ray points are provided by the BSDC database. (Bottom panel) The ligth-curve of Mkn 421, with data collected since 1980, showing the presence of variability in all timescales, from decades to days. The plots are made up of data from a dozen catalogues spanning more than 30 years, and collected by over 10 different instruments. The plots are produced with the SEDBuilder tool available online from ASDC.\protect\cite{Gio15} \label{fig1}}
\end{figure}

Blazars are a small sub-set of AGNs, distinguished by extreme observational properties, such as the presence of superluminal motion in high-resolution radio maps, and highly variable non thermal emission over the entire spectrum.\cite{Bla78} In blazars, such properties result from the fact that the plasma jets are pointing at a direction close to our line of sight, which amplifies the emission's relativistic effects.\cite{Urr95} Multi-frequency observations of blazars provide, therefore, a clear view of the physics and evolution of relativistic particles accelerated by the AGN. 

The spectral energy distribution (SED) of blazars displays two broad humps (see Fig.~\ref{fig1}). Radiation associated to the low-frequency part of the SED is firmly established as synchrotron emission from relativistic electrons interacting with the jet's magnetic field. The nature of the high-energy hump, peaking in the range from X- to gamma-rays, is attributed to two intrinsically different mechanisms: In a  purely leptonic scenario, emission is due to inverse-Compton scattering of soft photons by the energetic electrons;\cite{Sik94} whereas in lepto-hadronic models a component from proton-synchrotron radiation\cite{Muc01} or photo-hadronic interaction\cite{Pet15} may be present. 

The fact that a same population of relativistic particles is responsible for emission in the low and high-energy humps means that the entire SED of blazars is correlated, requiring joint multi-instrument data to be studied, over 20 orders of magnitude in energy. The kpc-scale sizes of the sources, contrasted with particle acceleration and cooling timescales as short as few hours or less, means that temporal analysis should span four orders of magnitude, from sub-hour to several years. Such observational characteristics clearly demonstrate the need for large integrated databases to provide the required data services which are beyond any individual group's capabilities to acquire and analyse. More recently, not only EM information such as shown in Fig.~\ref{fig1}, but also multi-messenger data, is becoming increasingly relevant in the field of blazar astrophysics, demanding, for example, software tools and data-services for cross-matching blazar catalogues and high-energy neutrinos and cosmic ray sky maps.\cite{Gio15}

Such "data-intensive" characteristics present in blazar studies are common to a number of fields within and beyond astrophysics, illustrating the global relevance of a model for an open access,  science and web-ready data center.

\section{The concept of the BSDC}

The BSDC is a space science data center which aims, primarily, to serve the interests and necessities of the Brazilian community of astronomy and astrophysics. It is a project under construction at the Brazilian Center for Physics Research (CBPF) which builds upon the experience and is being developed in close collaboration with ASDC, the science data center of the Italian Space Agency (ASI), where the concept of a science data center of this kind was originally advanced.\footnote{For a detailed presentation of the ASDC itself, please see the following document by Paolo Giommi\cite{Gio15} and visit the ASDC website at \url{www.asdc.asi.it}}. The BSDC is being built at CBPF, in Rio de Janeiro, within the framework of the International Center for Relativistic Astrophysics Network (ICRANet). It will be fully operational as an open access online data service in 2017. 

More than an online data repository, the BSDC will be a portal for multi-source data integration and access, and an online platform for research and education, containing software tools for data mining, visualisation, as well as data inspection and analysis. Not a simple data provider, its goal is to enable, promote and actively conduct collaborative data-intensive scientific research, whose success is measured in scientific impact and papers rather than Terabytes. It is focused therefore on astrophysics and space science applications characterised by "big data". The BSDC is built over two conceptual pillars, essential for its goals of integration and accessibility.

The fundamental notion behind the BSDC model is that of a "science-ready" database, that is, an on-line service providing access to scientific data in a form that can be directly used in professional scientific publications, without the need of any specific knowledge about the instrumentation that produced it. As in the case of blazars, data-intensive research requires information that is usually collected in complex ways, involving multiple instruments and techniques that no single group can individually master, or even guarantee its direct access to. Raw data repositories, even if open access, are thus limited in terms of direct scientific applications, requiring the knowledge of specialists to be analysed. Nevertheless, for most of the regular scientific applications, quality standard data products are enough, and providing direct access to such products cuts through an expensive and limiting intermediary to reaching the final scientific objective of the experiments. 

The purpose of the BSDC is therefore to provide final data products that can be directly used in scientific applications by a very broad fraction of the academic community, and easily integrated with other datasets from multiple sources. This allows for a truly global and democratic distribution and accessibility of the data, even by people or groups away from the centres where data is produced. By guaranteeing broad and easy integrability of related data from different origins, it also increases the scientific impact of individual datasets, specially from small and medium-sized facilities and observatories which in isolation cannot compete with world-class centres, but become relevant when integrated to those. 

The second conceptual pillar of the BSDC is that the database and all related tools must be "web-ready", that is, flexible and efficiently accessible for manipulation entirely over the web, guaranteeing robustness, stability and universality of the data services. This notion of web-readiness, combined with the numerous mobile platforms for web access today make the BSDC a real tool for "citizen science", meaning that it allows for first hand access to scientific data not only by scientists but to any interested citizen, as an expected return of society's investment in science. To this purpose the BSDC is fully supportive of the United Nations "Open Universe" initiative (see below for more detail).

The construction and operation of such a data center represents a prime opportunity for the formation and training of high-level human resources specialised in data technology, an essential skill today. It also has potential to serve as a center for development of applied data science research, at a scale likely unrivalled in Brazil.

\subsection{Current activities at BSDC}

Since the start of our activities in early 2016, the BSDC has already produced a few data products. Among them, we can cite access to two on-line astronomical catalogues, produced by Brazilian authors, on the fields of blazar astrophysics,\cite{Ars15} and white dwarf stars,\cite{Kep15} along with virtual observatory (VO) remote query services of these same tables. Discussions are ongoing with specific groups in Brazil for the future integration of software products and other
databases from National instruments and research within the BSDC, as soon as the center is fully operational in 2017. Likewise, the project has been presented at a BRICS Working Group Meeting on Astronomical Data\footnote{Please, see the contents of the 2nd BRICS Astronomy Workshop, held in Ekateringburg (Russia) in 2016, on the theme of "Astronomical Data and Computation" at {\url http://astro.brics.urfu.ru/en/astrodata2016/}.} on the context of a possible integrated expansion of the BSDC model to the BRICS community, given the relevant baseline of common necessities and interests of these countries in the field, as they are currently developing their national data infrastructure in astrophysics and space science.  

Since the BSDC is developed in cooperation with the ASDC and its data and software products can work over a common interface, specific functionalities and databases composing each center are readily available for the other. Likewise is the possibility for access with external resources like VO tools. In addition to products and services related to the Brazilian community, the BSDC shall also specialise in fields where expertise or infrastructure is not already in place within the ASDC and whose topic is of interest for national research groups. Two examples are the construction of an extensive database of ground-based very-high-energy gamma-ray astronomy data from multiple observatories around the globe (see example in Figure~\ref{fig1}), and a platform for optical polarimetric data, focused on blazar observations, both of which are services currently under development.

\section{The Open Universe Initiative}

The "Open Universe"\footnote{For more information, please refer to the proposal document for the "Open Universe" initiative, at \url{www.unoosa.org/oosa/en/oosadoc/data/documents/2016/aac.1052016crp/aac.1052016crp.6\_0.html}} is a recent initiative aimed at greatly expanding the availability and accessibility to space science data, extending the potential of scientific discovery to new participants in all parts of the world. A very wide range of communities will benefit from Open Universe: professional scientists, teachers and students, and potentially any citizen interested in space science. 

The initiative was proposed by Italy and presented to the United Nations Committee on the Peaceful Uses of Outer Space (COPUOS) in June 2016. The BSDC shares on the ethos of and supports the "Open Universe" initiative and in this context is also supported by the Brazilian Space Agency (AEB).

\section{Conclusions}

In this document we have presented the BSDC, a VO-compliant astrophysics database project under construction for Brazilian space science. It is developed in the spirit of creation of an international, open access, online data center aimed at enabling and actively conducting astrophysics and space-science data-intensive research through the integration of data bases and handling tools in a single, global, web platform.

The proposal of the BSDC is unique in its kind and extent in Brazil, developed with the support and cooperation with the ASDC, in a collaboration whereby it capitalises on decades of experience. It is also co-sponsor of a global initiative fostered under the auspices of the United Nations, which aims to bring together science and education oriented data centres into a worldwide framework to expand availability of and accessibility to open source science data.

The BSDC is open for collaboration with any interested group, specially those associated to Brazilian astrophysics and space science activities, and will be fully available for online access early in 2017 as a platform sharing on the database and functionalities of ASDC. Discussions are also being conducted in order to propose an expansion of the data center model for groups in other BRICS countries, through the ICRANet framework.

\section*{Acknowledgments}

The BSDC is partially funded by a FAPERJ Thematic Grant  110.148/2013. U.B.A. acknowledges a CNPq Productivity Research Grant (nr. 309606/2013-6) and a FAPERJ Young Scientist Fellowship (nr. 203.312/2016). B.B. was a recipient of a PCI-BEV, short-term Fellowship at CBPF, during the year 2016.

\end{document}